# Magnetoelectric Spin Wave Amplifier for Spin Wave Logic Circuits

[1)]Alexander Khitun, [2)]Dmitri E. Nikonov, and [1)]Kang L. Wang

[1)] Device Research Laboratory, Electrical Engineering Department,

Focus Center on Functional Engineered Nano Architectonics (FENA),

Western Institute of Nanoelectronics (WIN),

University of California at Los Angeles, Los Angeles, California, 90095-1594

[2)] Technology Strategy, Technology & Manufacturing Group, Intel Corporation, Santa Clara, California, 95054

Abstract

We propose and analyze a spin wave amplifier aimed to enhance the amplitude of the propagating spin wave via the magnetoelectric effect. The amplifier is a two-layer multiferroic structure, which comprises piezoelectric and ferromagnetic materials. By applying electric field to the piezoelectric layer, the stress is produced. In turn, the stress changes the direction of the easy axis in the ferromagnetic layer and the direction of the anisotropy field. The rotation frequency of the easy axis is the same as the frequency of the spin wave propagating through the ferromagnetic layer. As a result of this two-stage process, the amplitude of the spin wave can be amplified depending on the angle of the easy axis rotation. We present results of numerical simulations illustrating the operation of the proposed amplifier. According to numerical estimates, the amplitude of the spin wave signal can be increased by several orders of magnitude. The energy efficiency of the electric-to-magnetic power conversion is discussed. The proposed amplifier preserves the phase of the initial signal, which is important for application to logic circuits based on spin waves.

Introduction

Spin waves have been studied for many decades in a variety of magnetic materials and nanostructures [1-3]. Recent interest to spin waves is arisen from the intriguing possibility to use spin waves for logic devices [4]. In contrast to electron current whose spin polarization exists only over lengths of a few microns, spin wave can coherently propagate up to millimeter distances at room temperature, which makes them attractive for wave-like computing. The first working spin wave logic device utilizing spin wave interference has been experimentally demonstrated in room temperature [5]. However, spin waves have one significant shortcoming: the amplitude of the propagating spin wave exponentially decays due to the magnon-phonon, magnon-magnon and other scattering processes. Thus, the realization of the integrated spin wave-based circuits requires the introduction of a spin wave amplifier – a device aimed to provide gain in order to compensate losses during spin wave propagation. There are several well-known mechanisms, which can be used for spin wave amplification. For example, a spin wave can be amplified by passing electric current in a conducting ferromagnet along with the direction of spin wave propagation [6]. The use of an electric current for spin wave pumping may not be efficient from a power consumption point of view. A spin wave can also be amplified by an alternating magnetic field, via so called parametric parallel microwave pumping [7-10]. A microstrip structure can be used as a generator for ac magnetic field operating at the parametric resonance frequency $\omega = 2\omega_{sw}$, where $\omega_{sw}$ is the frequency of the spin wave signal. A parametric microwave spin wave amplifier having a gain coefficient up to 40 dB for the input power levels about 1 pW has been demonstrated experimentally [7]. However, the use of the microwave pumping has several technological disadvantages associated with direct coupling between the microstrips via the stray field. It may be beneficial form the practical point of view to use a local amplifier to restore spin wave amplitude at a certain point of the magnetic circuit.

In this work we describe a spin wave amplifier for spin wave amplitude enhancement using the magnetoelectirc coupling in a multiferroic structure. Multiferroics are a special type of material that possesses simultaneously electric and magnetic orders [11, 12]. This translates to the possibility to generate magnetic field by applying an electric field. There are only a few room temperature multiferroic materials known today [12], e.g. $BiFeO_3$ and its derivatives. An alternative method for obtaining a structure with magnetoelectric effect is a nanostructure consisting of two materials, a piezomagnetic film and a piezoelectric film [13]. The magnetoelectric coupling may arise as a combined effect of two: piezoelectricity and piezomagnetism. An electric field applied to the piezoelectric produces stress, which, in turn, affects the magnetic properties of the piezomagnetic film. The combined piezoelectric and piezomagnetic coupled material may be considered as a synthetic multiferroic structure, which causes the easy-axis rotation of the ferromagnet under the applied electric field. The angle of the easy-axis rotation depends on strength of the applied electric field. A two-layer magnetoelectric CoPd/PZT cell was experimentally demonstrated [14], and the easy axis rotation of up to 150 degree was observed. It was also experimentally demonstrated that a microwave planar resonator consisting of yttrium iron garnet (YIG) and ferroelectric barium strontium titanate (BST) thin films can be both electrically and magnetically tuned [15]..

In this paper, we consider a two-layer multiferroic structure consisting of a piezoelectric and a ferromagnetic materials. In Section II, we describe the material structure of the proposed spin wave amplifier and its principle of the operation. In Section III, we present results of numerical simulations illustrating the amplifier performance. In Section IV, we analyze the efficiency of electric-to-magnetic energy conversion. Finally, we summarize our results on the feasibility of magneto-electric spin wave amplifier.

II. Material Structure and the Principle of Operation

In Fig.1(a), we have schematically shown the material structure of the spin wave device with a magnetoelectric amplifier. From the bottom to the top, it consists of a semiconductor substrate (e.g. silicon), a conducting ferromagnetic film (e.g. CoFe), and a piezoelectric layer (e.g. PZT). The ferromagnetic film serves as a waveguide for spin waves. An external bias magnetic field $\vec{H}_b$ is applied along the X-axis. The dominant magnetization is ($M_x$) along the field, and so is the unperturbed direction of the easy axis. We focus on the spin waves, propagating perpendicular to the external magnetic field, usually called "Magnetostatic Surface Spin Waves" (MSSW). A bit of information is encoded into the *phase* of a propagating spin wave. Two initial phases 0 and π represent two logic states 1 and 0, respectively. The amplitude of the spin wave is subject to damping during propagation because of the scattering (e.g. magnon-magnon, magnon-phonon) and thus it exponentially decreases with the propagation distance. The aim of the amplifier is to increase the amplitude of the propagating spin wave while preserving the phase. We define the spin wave amplitude as the magnetization along the Z-axis: $M_z = A_0 e^{-\kappa y}\sin(\omega_{sw} t + \varphi)$, where $\kappa$ is the damping constant, and $A_0$ is the amplitude. The phase of the spin wave is the angle between the $M_z$ and $M_y$ projections. To illustrate the operation of the spin wave amplifier, in Fig.1(b) we depict the magnetization ($M_z$) in the wave propagating from the left to the right along the Y axis. The waveguide region of the ferromagnetic film covered by the piezoelectric has a length of $L$, between the coordinates $y=0$, and $y=L$ as shown in Fig.1(b). The amplitude of the spin wave decays ($\kappa>0$) everywhere in the ferromagnetic film except the area under the piezoelectric film ($0<y<L$). The amplitude of the spin wave is shown to increase as it propagates in the waveguide area under the piezoelectric layer. Hereafter, we will refer to this region as the amplification region.

The metallic contact on the top of the piezoelectric layer and the conducting ferromagnetic film (ground plane) serve as a set of two electrodes to apply voltage across the piezoelectric layer. If there is no voltage applied, we assume that the piezoelectric layer has no effect on spin wave propagation in the ferromagnetic layer. Under the applied bias $V_G$, the piezoelectric layer produces pressure, which changes the orientation of the easy-axis in the ferromagnet by the angle $\theta_G$. The stress produced by the piezoelement rotates the direction of the easy-axis in the Z-X plane, as it is shown in Fig.1(a), so the direction of spin wave propagation remains perpendicular to the magnetization of the ferromagnet. We vary voltage $V$ in accordance with $V = V_0 \sin(\omega_E t)$, where $0<V_0<V_G$, and the frequency of the ac electric field is equal the frequency of the spin wave signal $\omega_E = \omega_{sw}$.

The principle of operation is as follows. Initially, all spins under the piezoelectric layer are aligned along the X-axis. An incoming spin wave propagates from the left to the right along Y-axis as it is shown in Fig.1. The excitation of the input spin wave can be done by an outer device (e.g. microstrip), which is not discussed here. As the spin wave approaches the amplification region (the region under the piezoelectric), an electric voltage is applied to the top electrode. The applied electric field changes the direction of the easy axis in the X-Z plane (from the X axis toward the Z axis) via the magnetoelectric coupling. The change of the magnetic energy density caused by the easy axis rotation is equivalent to the appearance of an additional anisotropy field, which results in the magnetization precession around the effective magnetic field. The particular way of magnetization precession is defined by the small perturbation caused by the incoming spin wave ($M_z << M_s$). We want to stress that the *phase* (not the amplitude) of the incoming spin wave defines the final magnetization along or opposite to the new easy axis. The maximum magnetization projection on the Z-axis is a function of the rotation angle. In the ultimate limit of 90 degree rotation, when the new anisotropy field is along the Z-axis, the $M_z$ can be amplified to the saturation value of $M_s$. Next, when the applied voltage is decreased to zero, the easy axis returns to its initial state along the X-axis. The amplification mechanism is illustrated in Fig.2. In Fig.2(a), it is schematically depicted the evolution of the precession trajectory cased by rotation of the easy axis from the X axis toward the Z axis. The evolution is shown for two spin wave phases 0 and $\pi$, respectively. In both cases, the z-projection of the magnetization $M_z$ increases due to the z-component of the anisotropy field. Next, in Fig.2(b), it is shown the trajectory change while the easy axis returns to its original direction along the X axis. As we show in the next section by numerical modeling, the rotation of the easy axis with the same frequency as of the spin wave may result in the spin wave amplitude amplification depending on the rotation angle.

This principle of operation is different from the parametric amplification described in [8-10]. The difference is in the mechanism of the magnetic field generation. In the above cited works, the effective magnetic field is generated by the electric current in the outer contour (i.e. microstrip). In the proposed amplifier, the spin wave energy is pumped by the ac electric field via the magnetoelectric coupling. Some portion of the electric energy stored in the piezoelectric material can be converted to magnetic energy via the mechanical interaction between the piezoelectric and magnetostrictive materials. The amplifier is, in essence, a combined electric-mechanical-magnetic oscillator, providing a mechanism for conversion from electric to magnetic energy. The similar idea has been realized for voltage amplification using piezoelectric/piezomagnetic laminate composites. A magnetic-to-electric converter based on Terfenol-D/PZT laminates has been experimentally demonstrated [16]. Voltage output was generated by ac magnetic field via the magneto-electric coupling. A high voltage gain (more than 260) was observed. In our case, we consider the reverse mechanism (electric-to-magnetic conversion) to achieve a spin wave gain. A simplified analogy of the physical mechanism may be a gravity pendulum oscillating under a time varying gravitational force as it is shown in Fig.2(c). The length of the pendulum $l$ is equivalent to the magnetization $M$, while the time varying force of gravity $F$ is an analog to the effective magnetic field $H_{eff}$. The periodic oscillations of the gravity force may lead to the enhancement of the oscillation amplitude. The maximum amplitude is limited by the length of the pendulum $l$ (in our case, the maximum spin wave amplitude is limited by $M_s$).

III. Numerical Modeling

In order to evaluate the performance of the proposed amplifier, we present the results of its numerical modeling. The propagation of spin waves is described using the Landau-Lifshitz-Gilbert (LLG) equation [17, 18]:

$$\frac{d\vec{m}}{dt} = -\frac{\gamma}{1+\alpha^2} \vec{m} \times \left[ \vec{H}_{eff} + \alpha \vec{m} \times \vec{H}_{eff} \right], \tag{1}$$

where $\vec{m} = \vec{M}/M_s$ is the unit magnetization vector, $M_s$ is the saturation magnetization, $\gamma$ is the gyro-magnetic ratio, and $\alpha$ is the phenomenological Gilbert damping coefficient. The first term of equation (1) describes the precession of magnetization about the effective field and the second term describes its relaxation towards the direction of the field. The effective field is given as follows:

$$\vec{H}_{eff} = \vec{H}_d + \vec{H}_{ex} + \vec{H}_a + \vec{H}_b, \tag{2}$$

where $\vec{H}_d$ is the magnetostatic field ($\vec{H}_d = -\nabla \Phi$, $\nabla^2 \Phi = 4\pi M_s \nabla \cdot \vec{m}$), $\vec{H}_{ex}$ is the exchange field ($\vec{H}_{ex} = (2A/M_s)\nabla^2 \vec{m}$, $A$ is the exchange constant), $\vec{H}_a$ is the anisotropy field ($\vec{H}_a = (2K/M_s)(\vec{m} \cdot \vec{c})\vec{c}$, where $K$ is the uniaxial anisotropy constant, and $\vec{c}$ is the unit vector along the uniaxial direction $c \equiv (c_x, c_y, c_z)$), $\vec{H}_b$ is the external bias magnetic field. This formalism has been used to model the spin wave propagation in permalloy thin films and shown in good agreement with experimental data [2].

Hereafter, we apply LLG formalism to model local spin dynamics of the ferromagnetic film under the piezoelectric layer, assuming the input spin wave excited by the outer source (e.g. microstrip). The goal of our numerical modeling is to show the possibility of spin wave amplitude amplification by rotating the direction of the easy-axis $\vec{c}$. We further assume the easy-axis direction is driven by the electric field applied to the piezoelectric element and that it oscillates in phase with the applied ac bias. We assume that the easy-axis is along the X-axis, when no voltage is applied to the piezoelectric. When the ac voltage is applied, the easy-axis rotates in the X-Z plane, and the components of the unit vector are given by:

$$c_x = c_0 \sin\theta, c_z = c_0 \cos\theta, c_y = 0, \tag{3}$$

$$\theta = \frac{\pi}{2}\left(\frac{V}{V_G}\right),$$

where $\theta$ is the angle of the easy-axis rotation as a function of the applied voltage.

In Fig. 3(a), we show the results of numerical simulations illustrating the change of the normalized magnetization $M_z/M_s$ as a function of time for easy axis rotation with a maximum angle $\theta = \pi/4$. The dashed line depicts the input magnetization (magnetization change produced by the spin wave just before entering the region under PZT at $y=0$), and the solid line depicts the magnetization after amplification (magnetization change produced by the spin wave as it leaves the region under PZT at $y=L$). In our numerical simulations we used the following material parameters $\gamma$=2.0×10$^7$rad/s/Oe, $4\pi M_s$=10kG, $2K/M_s$=100Oe. The bias magnetic field is $H_b$=50Oe, and the Gilbert damping parameter $\alpha$=0.1. The gyro-magnetic ratio $\gamma$ and the saturation magnetization $M_s$ are taken for permalloy [2]. We intentionally choose an exaggerated anisotropy constant K (compared to permalloy), in order to illuminate the effect produced by the easy-axis rotation. The higher the anisotropy constant is, the larger the magnetization projection $M_z$ becomes. As seen from Fig. 3(a), the amplitude of the amplified wave may exceed the amplitude of the input wave by many times. For comparison, the typical damping time for spin waves in conducting ferromagnets (e.g. NiFe) is of the order of 1ns [2]. Taking into account typical MSSW group velocity of the order of $10^6$cm/s [19], the damping length can be estimated as 10 μm. Thus, the proposed amplifier may be used to compensate spin wave attenuation in ferromagnetic waveguides of the length of hundreds of microns at room temperature.

It is critical for spin wave amplifier to preserve the phase of the spin wave signal for use in a logic circuit as described in [4]. In Fig.3(b) we show the results of numerical simulations carried out for two cases when the incoming spin wave signal has two initial phases: 0 and π, respectively. As it is seen in Fig. 3(b), the output signals have the same amplitude, and the phase of the amplified spin wave retains the phase of the input wave. This result can be understood considering that the anisotropy energy is equal for magnetization along with and opposite to Z-axis, which correspond to the phases 0 and π.

The anisotropy field along the X-axis breaks the symmetry of the spin wave oscillations in the Z-Y plane. In Fig. 4, we show the trajectory of the in-plane (Z-Y) magnetization in the normalized coordinates $M_z/M_s$ and $M_y/M_s$, obtained as a result of numerical simulations for $\theta = \pi/4$ at $y=L$. The trajectory shows the magnetization evolution from the starting point ($M_x/M_s$ =1) in the time interval of $10\omega^{-1}$. The z-projection is larger than that of the y-projection, $M_z > M_y$, since the orientation along (or opposite to) the z-axis is more energetically preferable (anisotropy energy). The symmetry between z- and y-components is restored as the spin wave leaves the amplification region.

In Fig. 5, we summarize the results of numerical simulations showing spin wave amplitude amplification as a function of the angle of easy-axis rotation $\theta$ from 0 to $\pi/2$. There are three characteristic regions in this graph. First, when the rotation angle is relatively small, $\theta < \pi/20$, the energy obtained per cycle is not enough to compensate the damping during the propagation. As the angle reaches some critical value ($\theta = \pi/20$), spin wave amplitude remains the same, $A_{out} = A_{in}$. Second, beyond the critical angle, the amplitude of the output wave increases with the increase of the rotation angle. The spin wave amplitude gain may be significant depending the amplitude of the input spin wave. Third, the amplitude of the output wave saturates as $M_z$ becomes close to the $M_s$.

The presented results of numerical simulations were based on the assumptions with arbitrarily chosen parameters (K and α) and could be used only to illustrate the amplifier performance. From the results shown in Fig.3-5, we can project some important properties and general trends. However, in order to simulate realistic cases, taking specific material properties must be used.

IV. Energy efficiency

Conversion efficiency is the main parameter need to be estimated for the proposed amplifier. The described spin wave amplifier converts electric energy into spin wave energy via the magneto-electric coupling. Its figure of merit $\eta = \dfrac{\mathrm{E}_{sw}}{\mathrm{E}_{total}}$ is defined as the ratio of the energy delivered to the spin wave signal $\mathrm{E}_{sw}$ and the total energy consumed, $\mathrm{E}_{total}$. The total energy consumed can be expressed as follows:

$$\mathrm{E}_{total} = \mathrm{E}_{sw} + \mathrm{E}_{piezo} + \mathrm{E}_{Joule} \tag{4}$$

where $\mathrm{E}_{piezo}$ is the energy losses inside the electro-mechanical oscillator producing stress, and $\mathrm{E}_{Joule}$ is the energy dissipated due to Joule's heating in the conducting wires to bias the element.

These contributions to consumed energy can be estimated as follows:

$$\mathrm{E}_{sw} = \mu_0 S d_m H \Delta M \approx \mu_0 \chi S d_m H_a^2$$

$$\mathrm{E}_{piezo} = \frac{1}{Q_{me}} \frac{CV^2}{2} = \frac{\varepsilon_0 \varepsilon S d_e}{2 Q_{me}} E^2$$

$$\mathrm{E}_{Joule} = \frac{V^2}{R \omega_{sw}} = \frac{2\pi d_e^2}{R \omega_{sw}} E^2 \tag{5}$$

where $\Delta M$ is the magnetization change (spin wave amplitude) caused by the internal magnetic field $H_a$ due to the electro-magnetic coupling, $\chi$ is the magnetic susceptibility, $\varepsilon$ is the dielectric permittivity of the piezoelectric, $S$ is the area, and $d_m$ is the thickness of the ferromagnetic layer), $d_e$ is the thickness of the piezoelectric layer, $V$ is the voltage applied to the magnetoelectric element, $R$ is the wire resistance, $\omega_{sw}$ is the spin wave frequency, $Q_{me}$ is the quality factor of the piezoelectric oscillator and $E$ is the electric field across the metallic plates. The internal magnetic field $H_a$ and the electric field $E$ are related as follows:

$$H_a = \frac{1}{\alpha_{em}} E, \tag{6}$$

where $\alpha_{em}$ is the electromagnetic coupling coefficient.

In order to estimate the order of magnitude of the conversion efficiency, we use Eqs.(5) and the following material and structure parameters: $S$=100nm×100nm, $d_e$=100nm, $d_m$=100nm, $\omega_{sw}$=10GHz, $\varepsilon$=1000, $\chi$=800kAm$^{-1}$/100 Oe, $Q_{me}$ = 1, $R$ = 50Ω, and $\alpha_{em}$ = 50mV/(cm·Oe); the value of the electromagnetic coupling coefficient $\alpha_{em}$ is taken from the available experimental data [20]. The measured magnetoelectric constants vary from 42 mV/(cm·Oe) to 4,800 mV/(cm·Oe) for different two-phase multiferroic systems. In Fig. 6, we plotted the results showing each of the energies in Eq. (3) as functions of the bias electric field. As seen from Fig.6, most of energy is transferred to the spin waves, while only a small portion is dissipated inside the piezoelectric oscillator and Joule's heat. The losses in Joule's heat may exceed the losses in electro-mechanical oscillator depending on the contact wires resistance and the frequency of operation. The estimated figure of merit is about 97% for the voltage range under study. There are several comments regarding to this estimate. Our main assumption is that the magnetic energy is completely transferred into the spin wave. We neglected other relaxation processes (e.g. eddy currents), which would decrease the overall conversion efficiency.

At the same time, the quality factor $Q_{me}$ of the piezoelectric oscillator can be much higher than unity as reported in the experimental works [21, 22]. The conversion efficiency of magnetoelectric laminate structure described by S. Dong et al [16] was estimated to be 98% by neglecting eddy current. In general, the fundamental limit for the conversion efficiency of a parallel-plate capacitor filled by piezoelectric-piezomagnetic composites is defined by the ratio between the magnetic and electro-mechanical losses per amplification cycle, which can be derived from Eqs.(4-6) as follows:

$$\frac{\mathrm{E}_{mag}}{\mathrm{E}_{elect}} = \frac{\mu_0 \chi}{\varepsilon_0 \varepsilon} \frac{d_m}{d_e} Q_{em} \frac{1}{\alpha_{em}^2}. \tag{7}$$

Taking the same parameters we used for the numerical estimate shown in Fig.5, the ratio $\mathrm{E}_{mag}/\mathrm{E}_{elect}$ is about $10^5$. The amplifier efficiency for a given operating frequency can be optimized by the proper choice of the piezoelectric-piezomagnetic pair and geometry factors. For example, the electric energy stored and dissipated per circle can be decreased by scaling down the thickness of the piezoelectric material. The thinner is the thickness, the lower is the bias voltage $V$ required for anisotropy field generation. However, the minimum operation voltage in a real circuit is limited by the Johnson noise

$$V_{noise} = \sqrt{4k_B T Z B_L}, \tag{8}$$

where $k_B$ is the Boltzmann's constant, $Z$ is the circuit impedance at $f = 2\pi/\omega_{sw}$, $B_L$ is the circuit bandwidth. Taking $B_L$ =40GHz and Z=50Ω, we obtain the noise limit $V_{noise} \approx$ *0.2 mV*. In turn, the minimum operating voltage defines the maximum thickness of the piezoelectric layer (<1um) at which the spin wave amplification occurs, and, thus, defines the quality factors and the conversion efficiency. The latter leads us to the optimization problem in finding the set of material and geometry parameters providing the maximum conversion efficiency for a given signal frequency and an output amplitude. For example, our estimate in Fig.6 shows a 3000 *kT* output when the amplifier works at the Johnson noise limit. The output power may be enhanced with the same conversion efficiency by optimizing the structure geometry (scaling the thicknesses of the piezoelectric and ferromagnetic layers). However, the most important parameter is the electromagnetic coupling coefficient $\alpha_{em}$, which mainly defines the conversion efficiency of a given multiferroic structure.

A potential stability problem of the amplifier may come from the possible material structure imperfections. The variation of the ferromagnetic layer thickness may result in the variation of the anisotropy magnetic field along the length of the amplification region, which can produce an additional phase shift for the propagating spin wave. The sensitivity of the phase change with respect to the magnetic field change $\Delta\phi/\delta H$ can be estimated as follows [3]:

$$\frac{\Delta\phi}{\delta H} = -\frac{L}{d_m}\frac{\gamma^2(H+2\pi M_s)}{\omega^2 - \gamma^2 H(H+4\pi M_s)}. \tag{9}$$

The thinner the ferromagnetic layer is, the more sensitive the phase change becomes to the thickness variation. For example, the 10% variation of the 20nm thick NiFe ferromagnetic film may result in a $0.1\pi$ phase shift for 10 GHz spin wave per one micron propagation distance. Besides the undesirable phase shift, the variation of the material structure may reduce the amplification efficiency. There are other questions related to the feasibility of high-frequency (10-100GHz) magnetoelectric oscillators. It should be noted that the experiment values of the magnetoelectric coupling [20] are obtained for relatively low frequencies below 1kHz, and there is lack of high frequency (GHz) data. However, a high frequency (1GHz) PZT nanopowder based oscillators have been demonstrated [23]. To date, the experimental results show that the PZT material can be suitable for the applications of miniature RF devices with dielectric loss less than 0.02 in the RF region [23]. To the best of our knowledge, no high-frequency multiferroic oscillators have been realized so far. The lack of experimental data does not let us estimate the practically achievable quality factors. The realization of the spin wave amplifier requires the materials with high anisotropy. Because of the lack of data, we limit our scope to the describing the principle of operation of the proposed spin wave amplifier and the estimation its efficiency.

Conclusions

We described a spin wave amplifier employing magnetoelectric effect for spin wave amplification. The amplifier is an electromagnetic oscillator consisting of a two-layer multiferroic structure, which comprises piezoelectric and ferromagnetic materials. The energy of ac electric field is pumped into spin waves via the magnetoelectric coupling. The operation of the amplifier is illustrated by numerical modeling. The efficiency of the electric-to-spin wave energy conversion depends on the strength of the magnetoelectric coupling in a given multiferroic material and structure geometry. The estimated figure of merit is about 97% for a electromagnetic coupling coefficient of 50mV/(cm·Oe) taken from the available experimental data. The important feature of the proposed amplifier is the ability to preserve the phase of the spin wave signal. Potentially, magnetoelectric amplifiers can be used in spin-wave based logic circuits providing gain to compensate losses during spin wave signal propagation.

Acknowledgments

The work was supported in part by the Center of Functional Engineered Nano Architectonics (FENA) under the Focus Center Research Program (FCRP) and by the Western Institute of Nanoelectronics (WIN) under the Nanoelectronics Research Initiative (NRI).

Figure Captions

Fig.1. (a) Schematics of the spin wave amplifier. From the bottom to the top, it consists of a semiconductor substrate (e.g. silicon), a conducting ferromagnetic film (e.g. CoFe), and a piezoelectric layer (e.g. PZT). The ferromagnetic film serves as a waveguide for spin waves propagating perpendicular to the external magnetic field (MSSW mode). The amplitude of the propagating spin wave is amplified in the region under the piezoelectric layer via magnetoelectric coupling. The inset shows the direction of the easy axis rotation as a function of the voltage applied across the structure. (b) Spin wave amplitude ($M_z$) at the amplifier entrance *(y=0)*, and at the exit *(y=L)* of the amplifier. The amplitude of the spin wave is shown to increase as it propagates in the channel under the piezoelectric within *(0<y<L)*.

Fig.2. Illustration of the amplification process. (a) Evolution of the precession trajectory (blue curve) cased by rotation of the easy axis from the X axis toward the Z axis. (b) Precession trajectory while the easy axis returns to its original direction along the X axis. The evolution is shown for two spin wave phases 0 and π, respectively. (c) Spin wave amplitude enhancement as a result of the two-step process and a physical analogy: gravity pendulum under the time varying gravity force. The length of the pendulum *l* is equivalent to the magnetization *M*, while the time varying force of gravity *F* is an analog to the effective magnetic field $H_{eff}$.

Fig.3. (a) Results of numerical simulations illustrating the time evolution of the normalized magnetization $M_z/M_s$ of the easy axis rotation to the maximum angle $\theta = \pi/4$. The dashed line depicts the input magnetization at y=0, and the solid line depicts the magnetization after amplification at y=L. (b) The results of numerical simulations carried out for two cases when the incoming spin wave signal has two initial phases: 0 and π, respectively. The output waves preserve the phase of the input waves in both cases, while the amplitude for both phases are the same.

Fig.4. Trajectory of the in-plane (Z-Y) magnetization in normalized coordinates $M_z/M_s$ and $M_y/M_s$ from numerical simulations for $\theta = \pi/4$ at y=L. The trajectory shows the magnetization evolution from the starting point ($M_x/M_s$ =1) in the time interval of $10\omega^{-1}$. The trajectory is not symmetric with respect to z- and y-projections due to the change of the anisotropy field caused by the magnetoelectric coupling.

Fig. 5. Summary plot on spin wave amplitude amplification as a function of the angle of easy-axis rotation $\theta$ in the range from 0 to $\pi/2$. For the given material parameters, the amplification occurs at $\theta > \pi/20$. Beyond this critical angle, the amplitude of the output wave increases with the increase of the rotation angle. The amplitude of the output wave saturates as $M_z$ becomes close to the saturated magnetization, $M_s$.

Fig.6. Results of numerical simulations showing the energy of the output spin wave (blue curve), energy dissipated into Joule's heat (red curve), and the energy stored in the piezoelectric (orange curve) as a function of the applied electric field. The dashed line depicts the minimum spin wave energy at the Johnson noise limit ($V_{noise}$=0.2 mV,

$d_m$=100nm). The estimated figure of merit is about 97% in the considered electric field range.

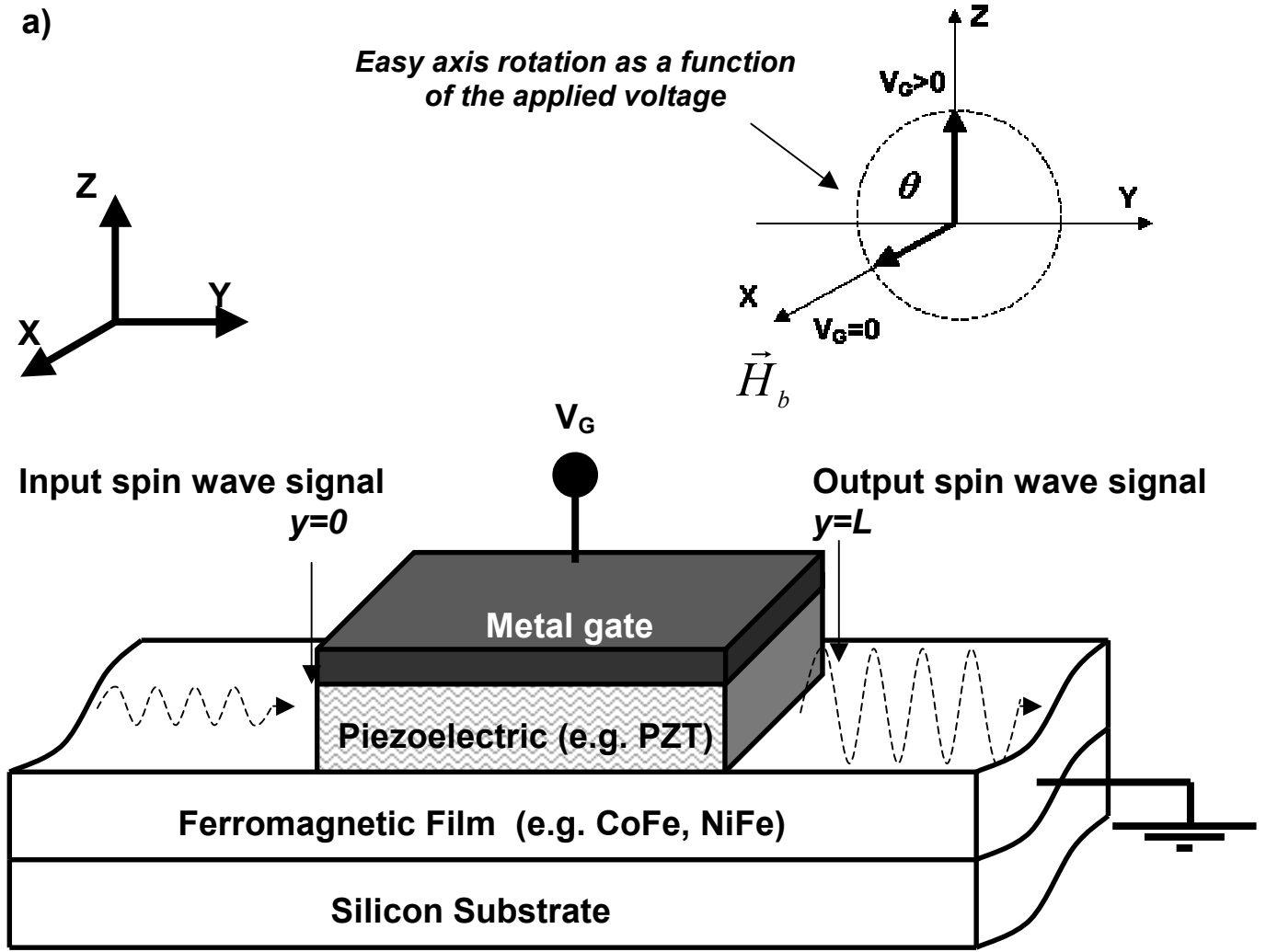

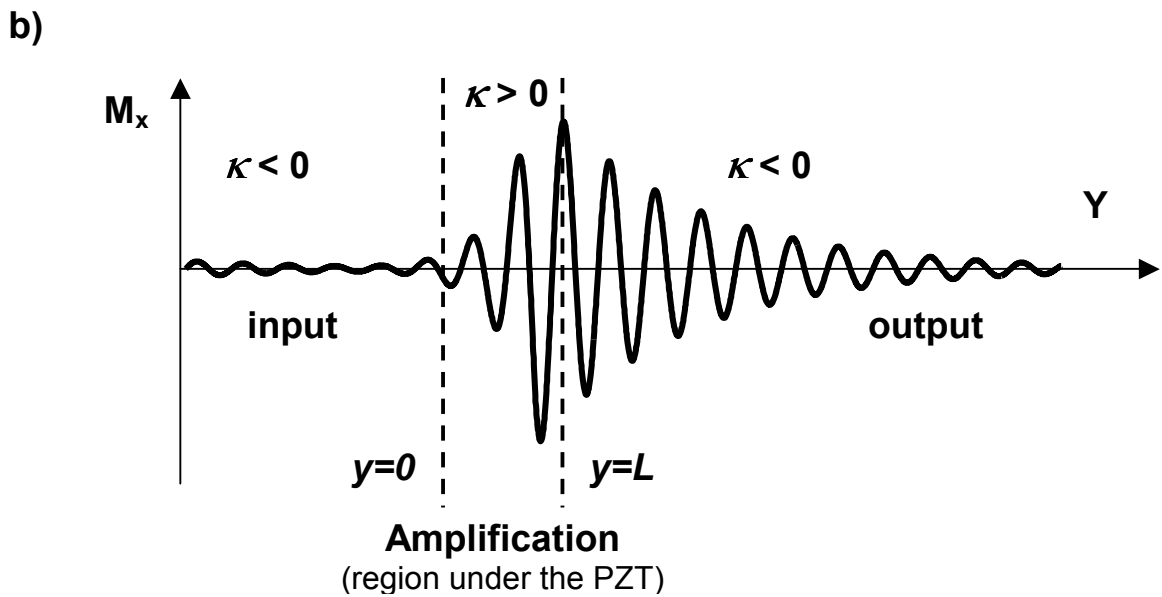

**Fig.1**

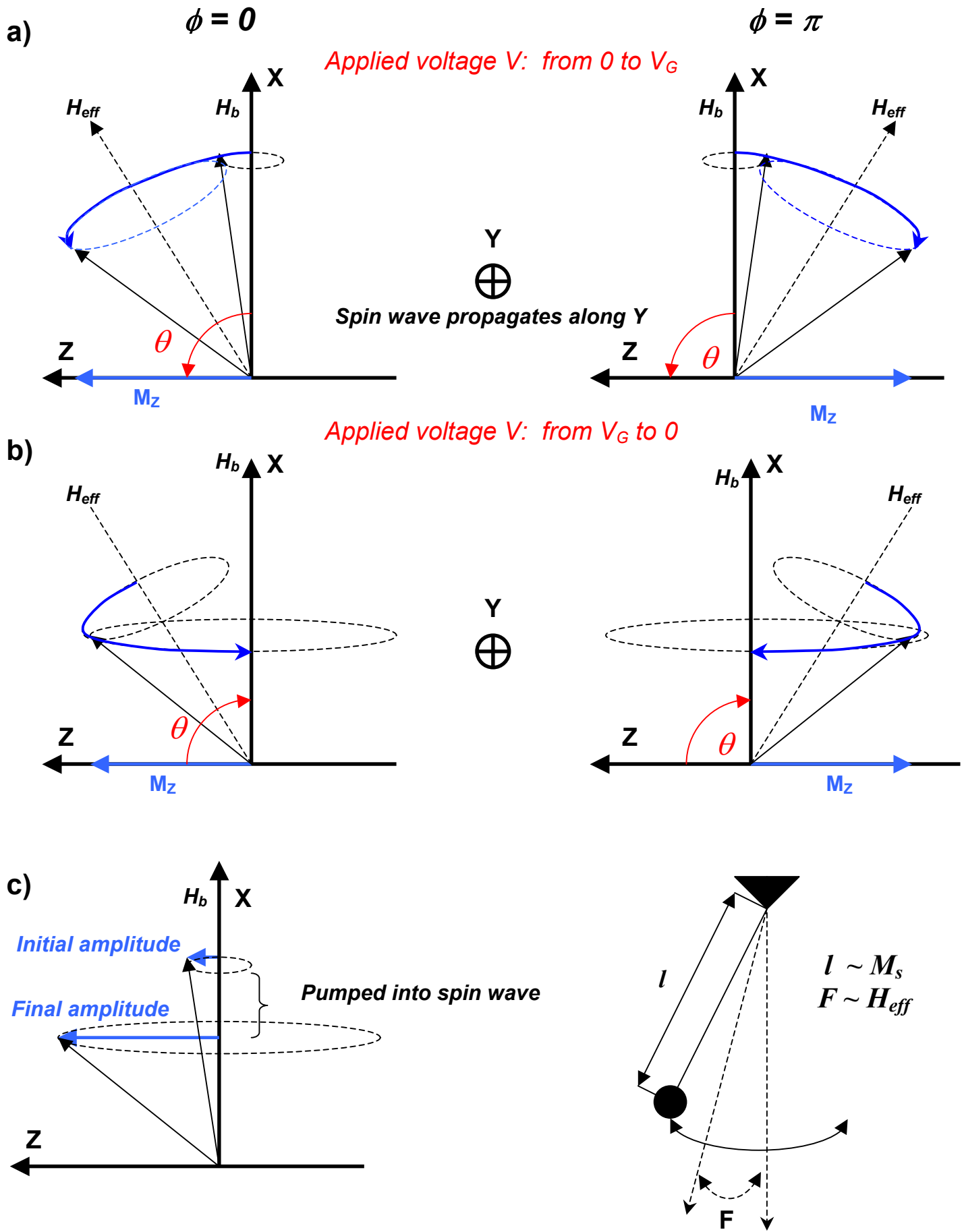

**Fig.2**

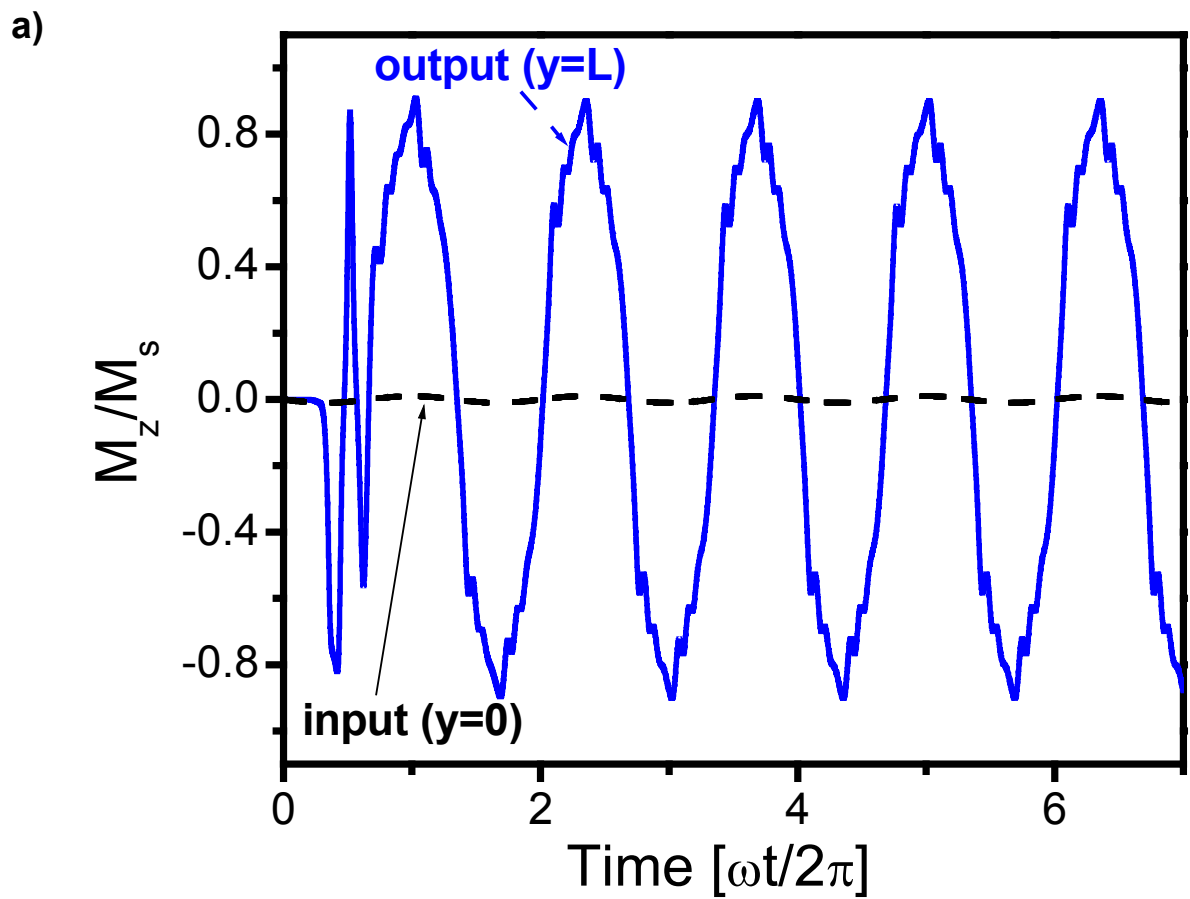

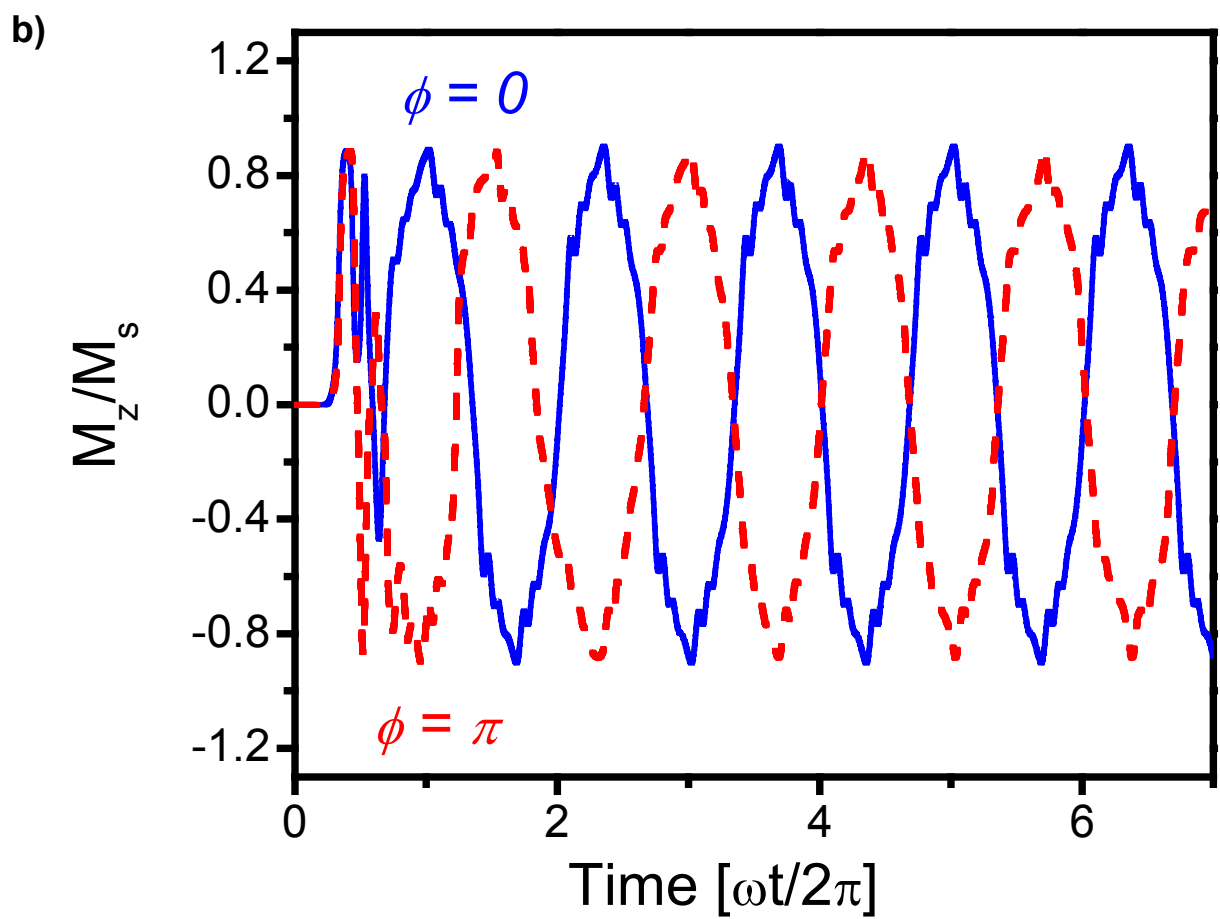

**Fig.3**

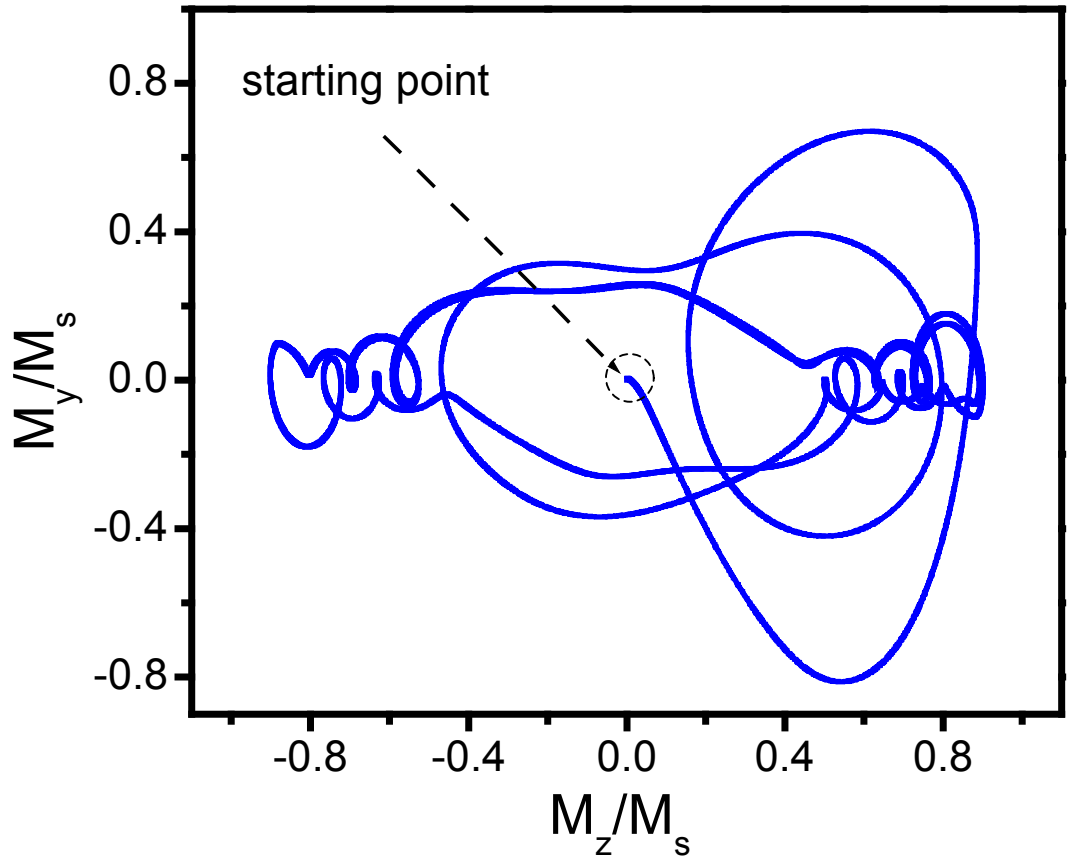

**Fig.4**

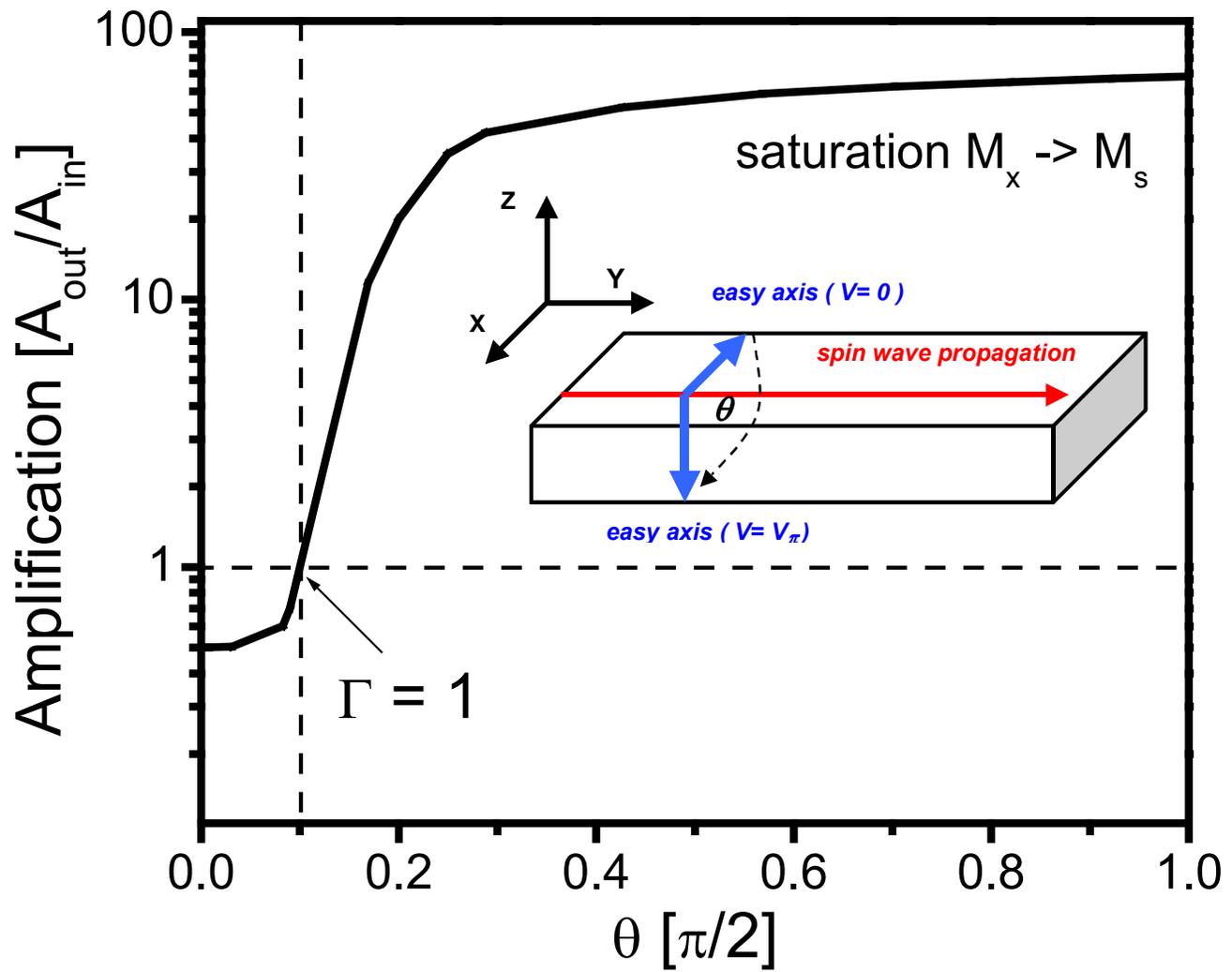

**Fig.5**

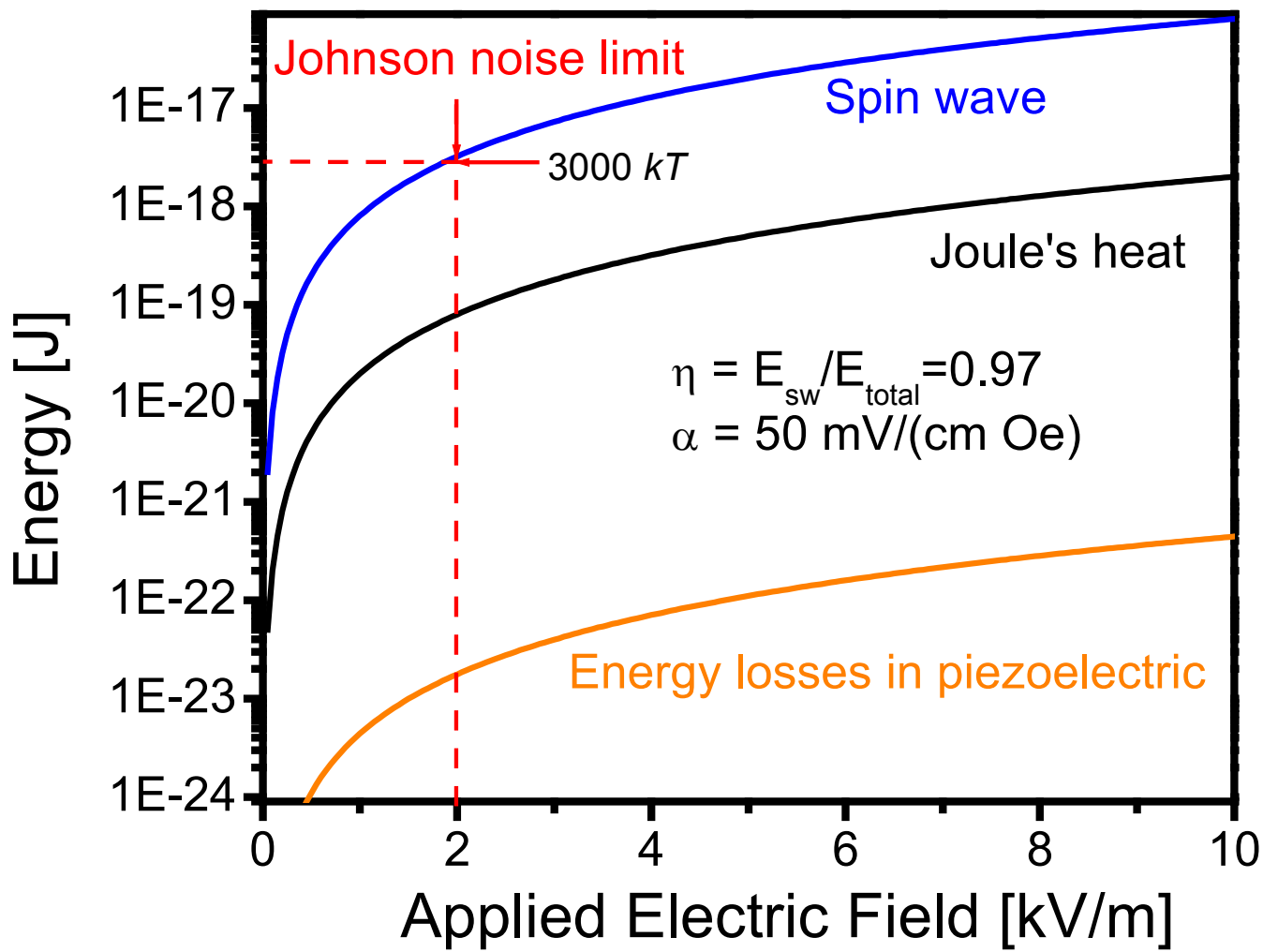

**Fig.6**